\documentclass[letterpaper]{article} 
\usepackage{aaai2026}  
\usepackage{times}  
\usepackage{helvet}  
\usepackage{courier}  
\usepackage[hyphens]{url}  
\usepackage{graphicx} 
\usepackage{natbib}  
\usepackage{caption} 
\usepackage{subcaption}
\usepackage{multirow}
\usepackage{algorithm}
\usepackage{algorithmic}
\usepackage{svg}
\usepackage{xcolor}         

\usepackage{newfloat}
\usepackage{listings}
\DeclareCaptionStyle{ruled}{labelfont=normalfont,labelsep=colon,strut=off} 
\floatstyle{ruled}
\newfloat{listing}{tb}{lst}{}
\floatname{listing}{Listing}
\graphicspath{figures}

\title{LabelKAN - Kolmogorov-Arnold Networks for Inter-Label Learning: \\ Avian Community Learning}

\author{
    Marc Grimson\textsuperscript{\rm 1},
    Joshua Fan\textsuperscript{\rm 1},
    Courtney L. Davis\textsuperscript{\rm 2},
    Dylan van Bramer\textsuperscript{\rm 2},
    Daniel Fink\textsuperscript{\rm 2},
    Carla P. Gomes\textsuperscript{\rm 1}
}
\affiliations{
    \textsuperscript{\rm 1}Cornell University Department of Computer Science\\
    \textsuperscript{\rm 2}Cornell University Lab of Ornithology\\
    mg2425@cornell.edu, jyf6@cornell.edu, cld74@cornell.edu, dcv26@cornell.edu, daniel.fink@cornell.edu, gomes@cs.cornell.edu
}

\begin{document}

\maketitle

\begin{abstract}

Global biodiversity loss is accelerating, prompting international efforts such as the Kunming-Montreal Global Biodiversity Framework (GBF) and the United Nations Sustainable Development Goals to direct resources toward halting species declines. A key challenge in achieving this goal is having access to robust methodologies to understand where species occur and how they relate to each other within broader ecological communities. Recent deep learning-based advances in joint species distribution modeling have shown improved predictive performance, but effectively incorporating community-level learning, taking into account species-species relationships in addition to species-environment relationships, remains an outstanding challenge. We introduce LabelKAN, a novel framework based on Kolmogorov-Arnold Networks (KANs) to learn inter-label connections from predictions of each label. When modeling avian species distributions, LabelKAN achieves  substantial gains in predictive performance across the vast majority of species. In particular, our method demonstrates strong improvements for rare and difficult-to-predict species, which are often the most important when setting biodiversity targets under frameworks like GBF. These performance gains also translate to more confident predictions of the species spatial patterns as well as more confident predictions of community structure. We illustrate how the LabelKAN leads to qualitative and quantitative improvements with a focused application on the Great Blue Heron, an emblematic species in freshwater ecosystems that has experienced significant population declines across the United States in recent years. Using the LabelKAN framework, we are able to identify communities and species in New York that will be most sensitive to further declines in Great Blue Heron populations. Our results underscore the critical importance of incorporating information on community assemblage in species distribution modeling. By leveraging species co-occurrence patterns, our approach offers deeper ecological insights and supports more informed conservation planning in the face of accelerating biodiversity loss. Beyond species distribution modeling, LabelKAN provides a principled approach to capturing inter-label connections and can generalize to diverse multi-label tasks. We hope it encourages further research on inter-label learning across domains.
\end{abstract}


\section{Introduction}

International commitments to reverse the global biodiversity crisis, including the targets outlined in the Kunming-Montreal Global Biodiversity Framework \cite{cbd} and the United Nations' Sustainable Development Goals \cite{sdgs}, highlight the urgent need for accurate, high-resolution information on species across large spatial extents to effectively guide conservation. Equally important is the need to understand co-occurrence patterns and species-species interactions, which are essential for capturing the structure and dynamics of ecological communities as a whole. Species distribution models (SDMs) are vital tools for predicting spatiotemporal patterns of species occurrence \cite{sdms}. While many SDMs focus on how these patterns are shaped by key environmental drivers such as climate, land cover, and land-use dynamics, there is strong ecological evidence that biotic interactions, such as competition, predation, facilitation, and other species-species relationships, are fundamental determinants of community structure and ecosystem function \cite{Wisz2012}. Incorporating these well-established drivers into models is critical, as their exclusion can lead to simplistic representations of ecological complexity. Recent advances in deep neural networks applied to SDMs \cite{dmvp,gmvae,deepmaxent,deepsdms} have significantly improved predictive performance but often fail to properly capture these important joint community dynamics, lack interpretable results for the label-label interactions, or only learn coarse interactions. For instance, deep models frequently over-predict common species at the expense of rare species \cite{deepsdms}, hindering their applicability to the ecological and conservation communities. This trade-off has slowed the broader adoption of deep SDMs, as conservation scientists rely not only on predictive accuracy but also on the ability to explain species' sensitivities to both environmental factors and community-level interactions—a necessary step for informing conservation plans and actions \cite{Guisan, eai_eco}.

To address these challenges, we propose an innovative framework, LabelKAN, which combines the strengths of neural networks for feature extraction and latent space construction with those of Kolmogorov-Arnold networks \cite{kans} for interpretability and improved predictive performance through learned label-label interaction patterns. We use a structured and interpretable latent space defined by an initial model as input into a series of KANs (our LabelKAN network). LabelKAN is then able to learn important label-label co-occurrence patterns that are otherwise missed, even in base models that are specifically designed to learn their own correlative structure. Using shallow KANs, the model learns smooth but highly informative connections between labels to improve species predictions at all levels of rarity. Our model is then easily analyzed for community-level understandings and co-occurrence patterns between species through consideration of the initial (potentially independent) community predictions. We view LabelKAN as a complementary module to existing approaches in multi-label classification, as it can be integrated atop state-of-the-art models to further refine predictions.

\textit{\textbf{Our contributions:}} \textbf{(1)} We introduce \textbf{LabelKAN}, a novel framework for improving inter-label learning that may be applied to any base model for multi-label classification by taking advantage of the unique strength of KANs over other networks. \textbf{(2)} We show through a series of experiments that LabelKAN improves state-of-the-art joint species distribution models for multi-label classification. \textbf{(3)} We analyze the resulting LabelKAN layers and effects to identify mechanisms for improving common and rare species. \textbf{(4)} Finally, through a careful analysis of the Great Blue Heron, a species of conservation concern, we demonstrate how the structured latent space learned by the LabelKAN can be used to infer novel conservation-relevant information delineating how further declines in this species are expected to change the structure of bird communities.

LabelKAN captures inter-label connections in a principled way and holds promise for a wide range of multi-label tasks. We hope this work will motivate further research on inter-label learning across domains.

\section{Related Works}

Multi-label classification: classic models include binary relevance \cite{binary_relevance}, which predicts each label independently of each other, possibly sharing earlier layers for extracting common high-level features, label power sets \cite{label_powerset}, which predict the power set of all labels, and classifier chains \cite{classifier_chains}, which turns the problem of multi-label classification into a series of classification tasks where the next classifier in the chain uses the results of the previous classifier. However, each of these has their own limitations. Binary relevance does not take into account any label relationships, which often results in poor performance on rare labels. The label power set transformation fails to scale to a large number of labels, as the power set grows exponentially. While classifier chains do take into account label relatedness, it requires a specific ordering of the labels where earlier classifiers in the chain do not see any information from later labels. This results in potentially very sensitive predictive performance highly dependent on the ordering of the labels. While there has been work on determining the best orderings \cite{annotationordermatters,cnnrnn}, one cannot get around the need for some type of ordering. To counter this, there have been a number of more recent works that aim to capture the label correlations and interdependence by capturing information from all labels at once. Deep Multivariate Probit model \cite{dmvp}, which extends the widely-used Multivariate Probit model to leverage deep learning for improved scalability in highly diverse species communities \cite{dmvp2}, explicitly learns a low-rank approximation of the correlation matrix. More recently, \cite{graphattention} uses a Graph Attention Transformer Network to learn the correlations between labels through an attention mechanism and a transformation of the label correlations to a graph structure. Our LabelKAN model is complementary to these models, as its construction allows for it to be placed on top of any existing multi-label classification framework, meaning any improvements to the base models result in further improvements from LabelKAN.

The idea of using a structured and interpretable latent space for enhanced learning and reasoning in models has been explored in other ares \cite{drnets}, but to our knowledge this is the first use of KANs on top of a structured and interpretable latent space representing label likelihoods.

Within the ecological SDM literature, there has been a surge of applications of deep learning methods, both for single-species and multi-species or joint species modeling approaches. Compared to traditional statistical modeling frameworks in ecology, such as Generalized Linear/Additive Models \cite{glmgam} or  hierarchical models \cite{sdmsimperfect}, deep learning offers enhanced scalability for large numbers of species, extensive observational data, and large feature sets. These models can often be separated into two groups, depending on the data used: presence-absence and presence-only. 

For presence-absence data, where we have both true positive and true negative labels for each location, there are a number of recent models. The DMVP model \cite{dmvp} was also initially applied to JSDMs, being one of the first models applied to ecology that considers the label correlations in a deep learning method, and still exists as a state-of-the-art method. Variational Autoencoder with Species Embeddings learned via contrastive learning, as originally proposed in \cite{gmvae}, uses a contrastive loss and shared latent space between label and feature embeddings to learn species-environment and implicit species-species relationships, and is another state-of-the-art JSDM method. 

In the case of presence-only data, however, we only have true positive labels but do not have information on what species were not present, which require different methodologies to model. The DeepMaxent model applies deep learning to maximum entropy modeling \cite{deepmaxent}, another common tool in statistical ecological modeling, though it does not take into account the label correlations. \cite{SINR} is a state-of-the-art method for presence-only data that introduces Spatial Implicit Neural Representations for species distribution models by learning spatial representations from the latitude and longitude for learning broad scale geospatial distributions. 

For a comprehensive review of deep learning applications in ecology, we refer readers to \cite{deepsdms}, though many of the models discussed focus on individual species predictions rather than explicitly modeling the joint distributions among multiple species in an ecological community.

\section{Joint Species Distribution Models}

A species distribution model (SDM) learns species-feature relationships by modeling the presence/absence of a single species as a function of features, $X_{lt}$, at a given location $l$ and time $t$, $l \in \mathcal{L}$ within a study extent $\mathcal{L}$ during some time $t \in T$ \cite{sdms,GUISAN2000147,Guisan}. Joint species distribution models (JSDMs) extend this approach by simultaneously modeling species-feature relationships for hundreds to thousands of species, $S$, to capture observed patterns of species co-occurrence across space and time \cite{hierarchicalsparse,jsdms,dmvp2}. Our goal is to learn a model, $\mathcal{M}_\theta$, parameterized by $\theta$ that predicts whether the species $s \in S$ is present $(Y_{lts} = 1)$, or absent $(Y_{lts} = 0)$ at location $l \in \mathcal{L}$ and time $t \in T$, based on the set of features $X_l$ and learned species-species associations.

\section{Kolmogorov-Arnold Networks}

\begin{figure}[t]
    \centering
    \begin{subfigure}{\linewidth}
        \centering
        \includegraphics[width=\linewidth]{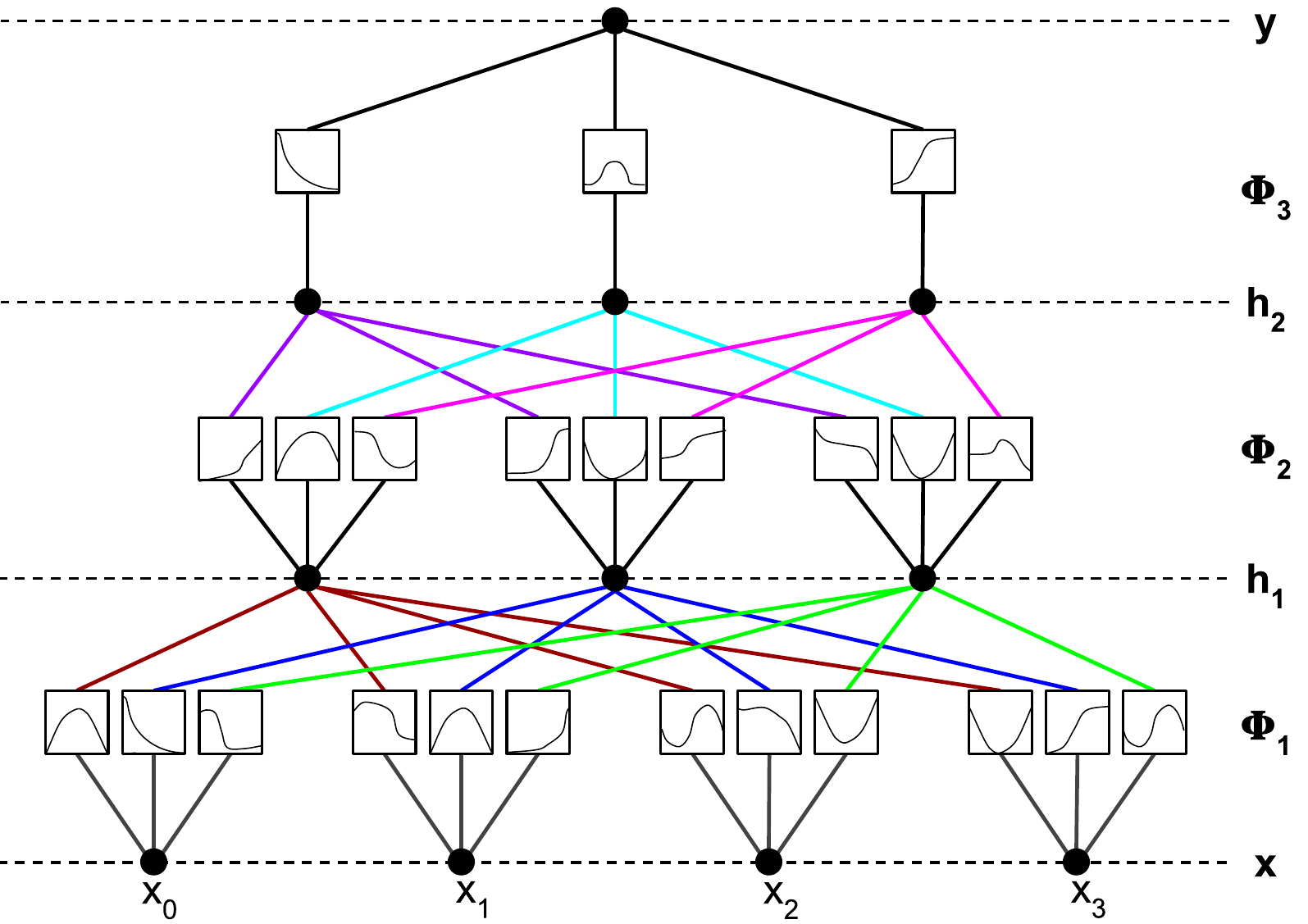}
        \caption{Example KAN with 4 inputs, two hidden layers of width 3, and a final single output.}
        \label{fig:kan}
    \end{subfigure}
    \begin{subfigure}{\linewidth}
        \centering
        \includegraphics[width=\linewidth]{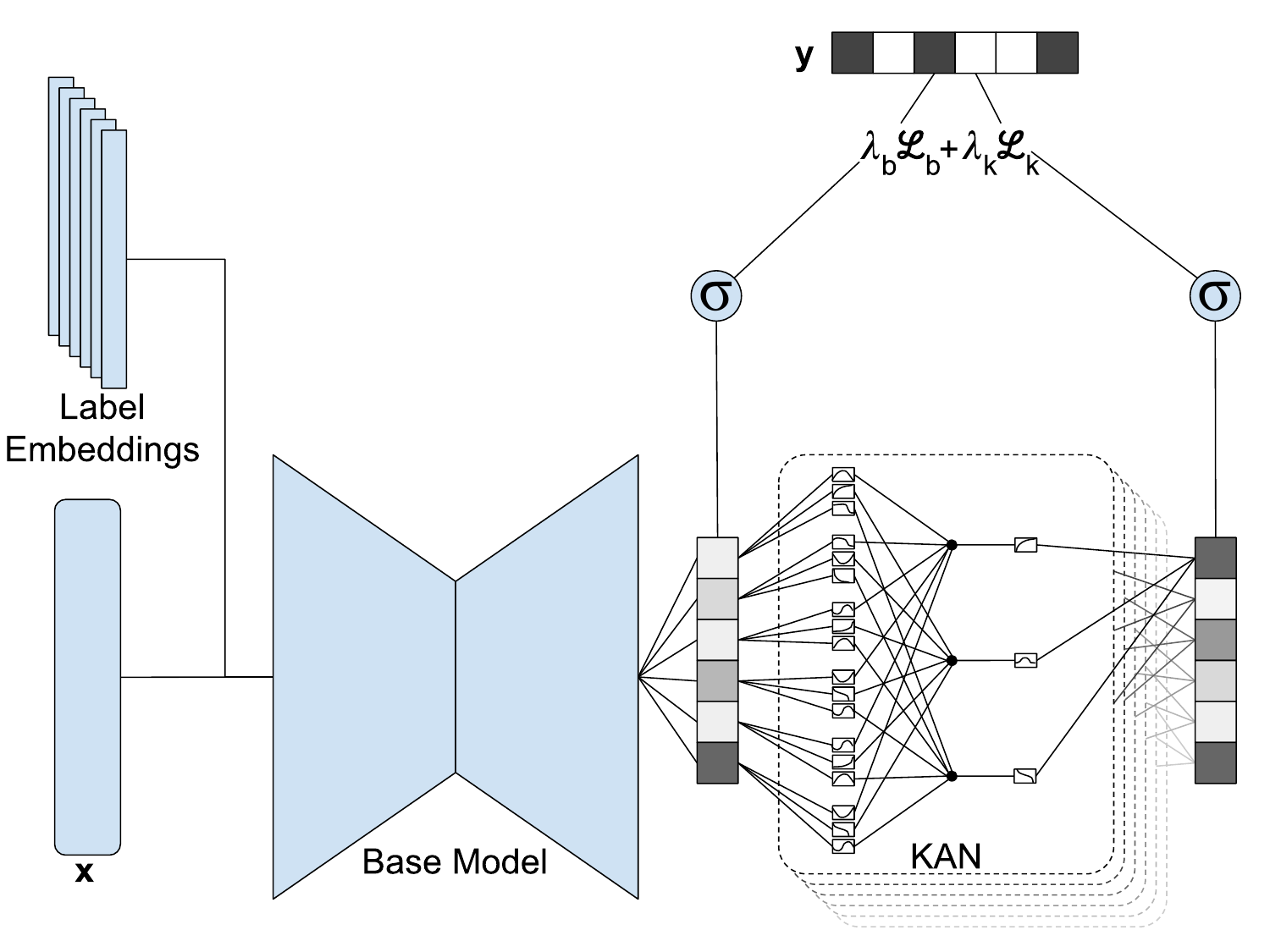}
        \caption{\textbf{LabelKAN} - features are passed through a base model to get initial predictions before being input into LabelKAN.}
        \label{fig:labelkan}
    \end{subfigure}
    \caption{(Top) Example 3-layer Kolmogorov-Arnold Network with learned activation functions. (Bottom) LabelKAN model abstraction with KAN heads.}
\end{figure}

Kolmogorov-Arnold Networks (KANs) \cite{kans} are a new deep learning model based on the Kolmogorov-Arnold Representation Theorem, as opposed to the Universal Approximation Theorem for standard neural networks. Instead of applying an activation function, such as ReLU, to the weighted sum of the outputs of the previous layer, KANs construct a univariate function of each input (calculated using splines) and take the weighted sum of the univariate functions. An example KAN is shown in Figure \ref{fig:kan}. Thus, there is no single activation function, but instead a learned activation function via the univariate spline. By stacking these KAN layers, KANs can theoretically approximate any function \cite{kans}, just like neural networks. While KANs tend to be slower to train, they also tend to be more compact and interpretable than neural networks, due in part to their smaller size, as well as the ability to visualize the splines and summations applied to each input feature.

\section{LabelKAN}

We propose a novel application of KANs to learn label co-occurrence patterns and capture how the likelihood of other labels, conditioned on the initial features, influences each label's final prediction. Consider  an initial model $\mathcal{M}$ that takes as input a set of features $X$ and predicts a probability of each of the $m$ labels occurring:

\begin{equation}
    y_\mathcal{M} = \sigma(\mathcal{M}(X)) \in [0,1]^m
\end{equation}
where $\sigma$ is the sigmoid function to enforce $y \in [0,1]^m$ and $m$ is the number of labels. We  attach a KAN head $\mathcal{K}_i$ for each label $i$ to predict that takes $\mathcal{M}(X)$ as input and outputs a final prediction,
\begin{equation}
    y_{\mathcal{K}_i} = \sigma(\mathcal{K}_i(\mathcal{M}(X))) \in [0,1]
\end{equation}
for each label $i \in [m]$, as shown in Figure \ref{fig:labelkan}. 

Although a single KAN could be used to output $y_\mathcal{K} \in [0,1]^m$, similar to the base model, we find that training a separate KAN head for each label provides better interpretability and simplifies training. A key strength of this structure is its flexibility: the underlying base model is interchangeable, as we demonstrate in our experiments, as long as it produces a prediction  for each label. While scaling this approach to a large number of labels may be challenging, each KAN head is parallelizable on a GPU. In practice, the approach scales effectively to the 200-label setting considered in our work.

To help ensure the separation of the label-feature and label-label signals, we apply a binary cross-entropy loss to the sigmoid outputs of both the base model and the LabelKAN. By maintaining this separation, both models are encouraged to capture domain-relevant interpretable signals. Without this constraint, the latent space may lack meaningful structure and interpretability, and prevent the LabelKAN from effectively learning inter-label relationships. 
Our loss function includes both binary cross-entropy terms: 

\begin{equation}
    \mathcal{L} = \lambda_\mathcal{M}\mathcal{L}_{BCE}(y_\mathcal{M}, y) + \lambda_\mathcal{K}\frac{1}{m}\sum_{i=i}^m\mathcal{L}_{BCE}(y_{\mathcal{K}_i}, y_i)
\end{equation}

For our tests, we choose $\lambda_{K} = 2 \lambda_{M}$.

\section{Experiments}

We run experiments for both presence-absence (PA) and presence-only (PO) datasets. Both runs focus on the 200 most frequently detected species in New York state during the breeding season, representing a taxonomically diverse group that includes both migratory and resident species.

For all experiments, we train both a set of base models and for each base model, we use the pretrained model as a warm start and then train the entire model end-to-end with LabelKAN, which tends to train faster than end-to-end training from the beginning. Results of models trained fully from scratch and with frozen base model parameters are provided in the supplement. All LabelKAN networks are shallow networks with the $200$ species logits as input, a single hidden layer of size $8$, and a final output layer of size $1$, with one such KAN per species.

\subsection{Presence-Absence}

We use species observational data from the eBird participatory science program\cite{ebird} collected during the breeding season (May 1 - August 2) from 2014 to 2019 within New York state. This dataset consists of "complete" species checklists, in which observers report all bird species detected at a given location and time, which are our labels. The 200 selected species span a range of prevalence and geographic distribution, from common (max $56.66\%$ of checklists) to infrequently detected species (min $0.22\%$ of checklists). From these species observations, we constructed a label vector describing the detection/non-detection of the 200 species for the \textit{n} = 125,389 complete checklists.

To estimate the joint occurrence of these species, we incorporate 102 features representing four key aspects of the data generation process: (1) 6 \textbf{observation process features} to account for variation in detection rates due to survey effort, observer expertise, and other observational biases; (2) 30 \textbf{climate features} \cite{daymet} capturing the effects of temperature and precipitation, including both long- and short-term means and variability; (3) 6 \textbf{topography features} \cite{topo1,topo2} to account for local geomorphology such as elevation and landform characteristics; and (4) 60 remote sensing \textbf{land cover features} \cite{modis} to provide information on habitat structure and vegetation characteristics. Additional details are provided in the supplementary material.

All models are trained for a minimum of 50 epochs and up to 200 epochs on a single NVIDIA \textregistered Tesla \textregistered V100 using PyTorch \cite{pytorch} and AdamW optimizer \cite{adamw} with default learning rate and beta parameters, which were selected through hyperparameter sweeps. Model sizes were chosen to balance performance and training time, with similar improvements across different sizes. The best-performing model is selected based on the epoch with the lowest validation loss, with early stopping triggered after 10 consecutive epochs without improvement.
\textit{}
We train three base models, including two state-of-the-art JSDMs with publicly available code.: (1) NN, a simple feedforward neural network; (2) DMVP, the Deep Multivariate Probit model \cite{dmvp,dmvp2}; and (3) VAE, a Variational Autoencoder with contrastive learning to capture label correlations \cite{gmvae}. The NN consists of five hidden layers (sizes [128, 256, 256, 128, 64]) using the ReLU activation function and a dropout of 0.05. Input and output sizes corresponding to the number of input features and species, respectively. The DMVP consists of three hidden layers (sizes [128, 256, 128]) and a latent dimension of 32 for the low-rank correlation matrix, also with ReLU activation functions and a dropout of 0.05. The VAE consists of an encoder and decoder, each with three hidden layers (sizes [128, 256, 128]), a latent dimension of 32, and a species embeddings space of size 128, also with ReLU activations and dropout of 0.5 applied throughout.

For each model we calculate both micro-aggregated and macro-aggregated statistics for the area under the precision-recall curve (AUPRC) and area under the receiver-operator curve (AUROC), where micro-aggregation computes metrics by pooling all predictions into a single class and macro-aggregation computes the metrics across each class separately and then takes the mean. We chose these metrics as they are able to represent results across all labels and within labels, with both metrics being important for consideration within class imbalanced datasets \cite{auprcauroc}.

The results in Table \ref{tab:labelkan} show that LabelKAN improves performance across all metrics compared to the respective base models alone, regardless of the base model, with the VAE + LabelKAN having the best performance.

\begin{table}[t]
    \centering
    \begin{tabular}{l|c|c|c|c}
        \multirow{2}{*}{Model} & \multicolumn{2}{c}{AUPRC} & \multicolumn{2}{c}{AUROC} \\
        & Micro & Macro & Micro & Macro \\
        \hline
        NN & 0.600 & 0.408 & 0.944 & 0.913 \\
        DMVP & 0.603 & 0.410 & 0.948 & 0.914 \\
        VAE & 0.612 & 0.421 & 0.949 & 0.916 \\
        NN-KAN & 0.624 & 0.434 & 0.951 & 0.919 \\
        DMVP-KAN & 0.624 & 0.436 & 0.951 & 0.919 \\
        VAE-KAN & \textbf{0.631} & \textbf{0.446} & \textbf{0.952} & \textbf{0.921} \\
    \end{tabular}
    \caption{Results of different base models (NN - neural network, DMVP - deep multivariate probit, VAE - variational autoencoder) with and without the LabelKAN applied (-KAN). The LabelKAN improves over all respective base models, with the VAE-LabelKAN performing the best across all metrics.}
    \label{tab:labelkan}
\end{table}

\subsection{Presence-Only}

\begin{table}[t]
    \centering
    \begin{tabular}{l|c}
        Model & Mean Average Precision \\
        \hline
        SINR & 0.494 \\
        SINR + LabelKAN & \textbf{0.534} \\
    \end{tabular}
    \caption{Results of the Spatial Implicit Neural Representation (SINR) model on presence-only data with and without LabelKAN, trained on iNaturalist data and tested on eBird S\&T data.}
    \label{tab:sinrkan}
\end{table}

For the presence-only dataset, we use iNaturalist \cite{inat} data from the SINR paper \cite{SINR}, subset to just the 200 New York state species. For this experiment we just use the longitude and latitude coordinates as features, passed through an encoder that generates $\sin$ and $\cos$ of the coordinates. We use the full "assume negative" loss proposed in \cite{SINR} to train the base SINR model with and without LabelKAN. All models are trained for a minimum of 10 epochs, up to 50 epochs, on a single NVIDIA \textregistered Tesla \textregistered H100 using PyTorch \cite{pytorch} and AdamW optimizer \cite{adamw}. For each model we calculate the mean average precision, as was done in the original SINR paper, with results shown in Table \ref{tab:sinrkan} on just the eBird status and trends dataset provided by \cite{SINR}, subset to the 200 species\footnote{The S\&T dataset only had 189 shared species, so metrics were calculated on the shared species, though all 200 species were trained and predicted}. Just as with the presence-absence data, we see an improvement with the addition of LabelKAN.

\subsection{Ablation Study}

To determine whether the observed performance gains are due to the extra KAN layers, or whether similar improvement could be achieved by applying any model for inter-label learning, we conduct an additional experiment. Using the same base models as before, we replace the LabelKAN with a neural network to do the label learning, effectively creating a deeper architecture while retaining the structured latent space. As shown in Table \ref{tab:labellearners}, replacing the KAN with another model for label learning yields no significant performance gain over the base model, suggesting that the benefits are specific to the KAN architecture.

\begin{table}[t]
    \centering
    \begin{tabular}{l|c|c|c|c}
        \multirow{2}{*}{Model} & \multicolumn{2}{c}{AUPRC} & \multicolumn{2}{c}{AUROC} \\
        & Micro & Macro & Micro & Macro \\
        \hline
        NN & 0.617 & 0.414 & 0.948 & 0.903\\
        LabelKAN & \textbf{0.631} & \textbf{0.446} & \textbf{0.952} & \textbf{0.921}
    \end{tabular}
    \caption{Results of using the same base model (VAE) but changing out the LabelKAN for a standard deep neural network. The neural network yields negligible improvement, whereas LabelKAN achieves substantial performance gains.}
    \label{tab:labellearners}
\end{table}

\section{Analyses}

Moving beyond aggregate predictive performance, we analyze the effects on individual species. First, we examine how predictive performance and learned patterns by LabelKAN vary with species rarity. Then, we conduct a detailed case study of the Great Blue Heron, an emblematic species experiencing significant decline across New York state \cite{nadecline,ebirdsnt}.

\subsection{Rare Species}

Considering species co-occurrence may particularly benefit rare species, as information from co-occurring species may improve rare species predictions. Here, we examine the effects of LabelKAN on rare versus common species.

\begin{figure}[t!]
    \centering
    \begin{subfigure}{\linewidth}
        \includegraphics[width=\textwidth]{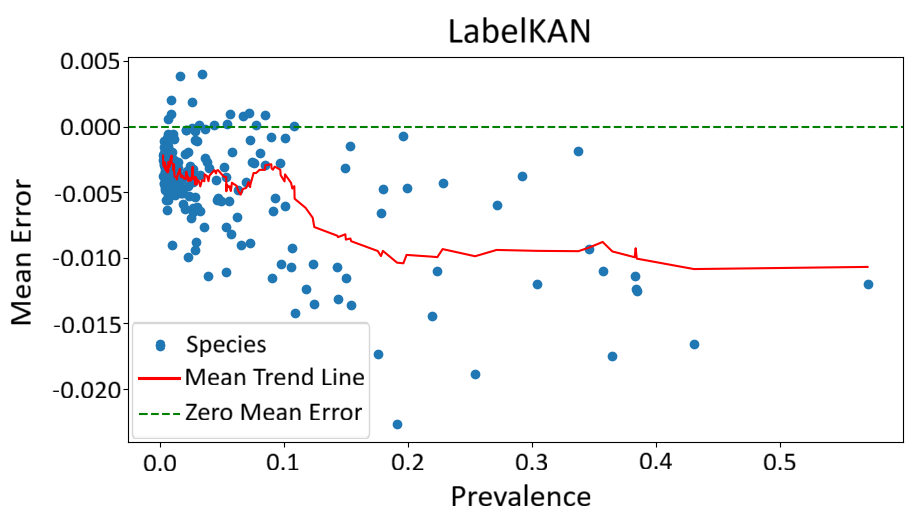}
        \caption{LabelKAN error vs. species prevalence}
        \label{fig:kanerror}
    \end{subfigure}
    \begin{subfigure}{\linewidth}
        \includegraphics[width=\textwidth]{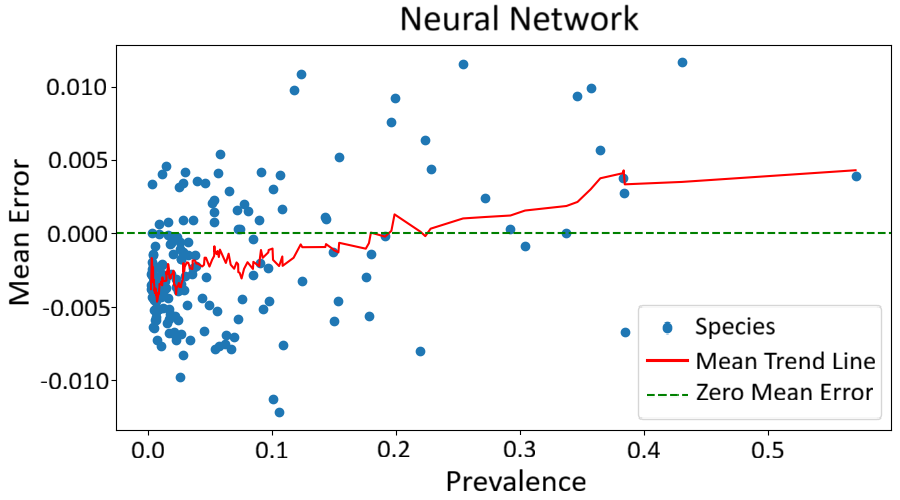}
        \caption{Neural Network label learner error vs. species prevalence}
        \label{fig:nnerror}
    \end{subfigure}
    \caption{Comparison between LabelKAN (top) and a Neural Network label layer (bottom) as a function of the mean change in error versus the species prevalence. On average, the LabelKAN model performs better at all prevalence levels over its base model, with only $7.5\%$ of species performing worse, with the worst performance reduction of only $0.4\%$. In contrast, for the Neural Network, nearly $25\%$ of species perform worse than the base line model, with a worst performing species degradation of $1.16\%$ average error.}
    \label{fig:errors}
\end{figure}

Given a LabelKAN model, we have predictions from both the base model $\mathcal{M}$ and the LabelKAN layers $\mathcal{K}_i$. For each species $i < n$, we consider the true label $y^i$ and the predicted values $\sigma(\mathcal{M}(x)_i)$ and $\sigma(\mathcal{K}_i(x))$ from the base model and LabelKAN, respectively, given features $x$. We compute the mean absolute error in predictions across the test set $X, Y \in \mathcal{T}$ to get $e_\mathcal{M}^i = \frac{1}{|\mathcal{T}|}\sum_{x, y \in \mathcal{T}}|y^i - \sigma(\mathcal{M}(x)_i)|$ and $e_\mathcal{K}^i = \frac{1}{|\mathcal{T}|}\sum_{x, y \in \mathcal{T}}|y^i - \sigma(\mathcal{K}_i(x))|$. We then analyze the difference $e_\mathcal{M}^i - e_\mathcal{K}^i$, where positive values indicate that the base model, on average, outperforms LabelKAN for species $i$, and negative values indicate improved performance from LabelKAN. Figure \ref{fig:kanerror} shows these results for the best performing model, LabelKAN, and its base model, the VAE, as a function of species prevalence, the average species occurrence rate across the study area. While $7.5\%$ of species have reduced performance, most rare species do benefit from LabelKAN. More common species show notably improved performance. In contrast, when replacing LabelKAN with a standard Deep Neural Network for label learning (Figure \ref{fig:nnerror}), we see that while some species improve, this often comes at the expense of many other species, resulting in minimal overall improvement. In the next section, we explore the potential reasons for performance gains in rare versus common species and show that the mechanisms behind these improvements differ.

\subsection{Learned Patterns}

In this section, we examine the patterns learned by LabelKAN models. Due to the use of KANs, it is possible to visualize and interpret the functional form and effects of individual input features - in this case, species logit values - by inspecting the learned univariate functions and the strength of each connection, as described in the original KAN papers \cite{kans,kan2}. As discussed in the previous section, LabelKAN improves performances for both rare and common species, but with more substantial error reduction for common species. We propose that these improvements arise from two different mechanisms: for common species, performance gains result from increased confidence in the base model's initial predictions; for rare species, improvements stem from leveraging information from other, co-occurring species in conjunction with its own initial prediction.

\begin{figure}[t]
    \centering
    \includegraphics[width=\linewidth]{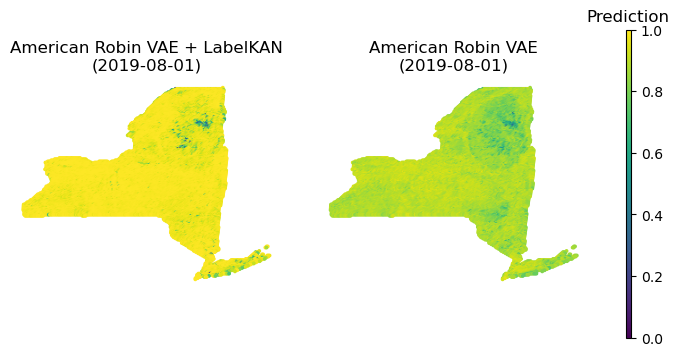}
    \caption{Predictions for the American Robin, a common species, on August 1, 2019. LabelKAN (left) results in more confident predictions (closer to the 0/1 value) across the spatial extent than the base VAE (right).}
    \label{fig:amerob}
\end{figure}

\begin{figure}[t!]
    \centering
    \begin{subfigure}{\linewidth}
        \centering
        \includegraphics[width=0.8\linewidth]{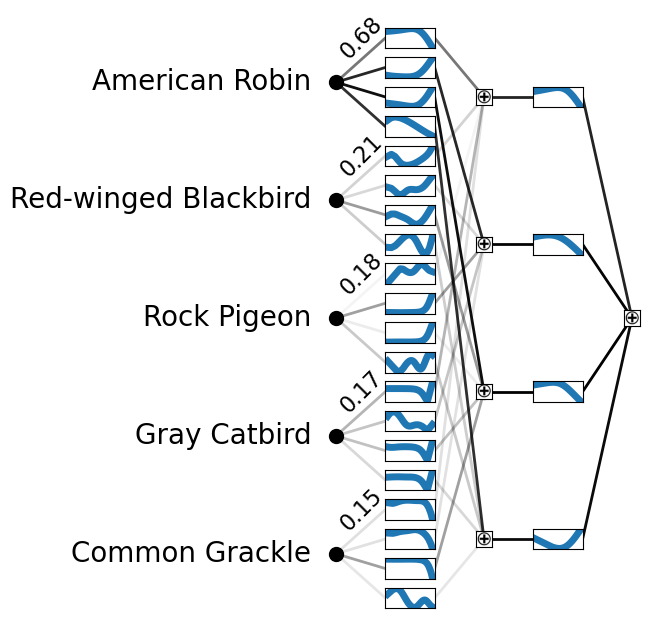}
        \caption{American Robin}
        \label{fig:amerob_kan_plot}
    \end{subfigure}
    \begin{subfigure}{\linewidth}
        \centering
        \includegraphics[width=0.8\linewidth]{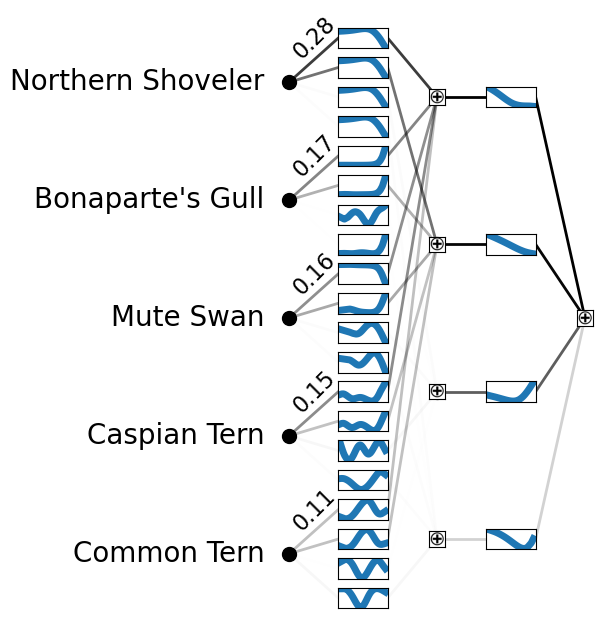}
        \caption{Northern Shoveler}
        \label{fig:norsho_kan_plot}
    \end{subfigure}
    \caption{Subset of the learned functions and edge connections from the LabelKAN model for the American Robin (top) and Northern Shoveler (bottom), the most and least prevalent species respectively, showing the top 5 contributing species. Each plot has, for the first layer, the average relative strength \cite{kans}. For the American Robin, most of the signal comes from itself, whereas for the Northern Shoveler, the signal strength from the input species logits is more distributed across co-occurring species.}
    \label{fig:species_kan}
\end{figure}

Take the American Robin as an example - the most prevalent species in the study region, appearing in $56.66\%$ of eBird checklists for New York. As shown in Figure \ref{fig:amerob}, the predicted occurrence maps from both the base model and LabelKAN show similar spatial patterns. However, the LabelKAN predictions are more confident (closer to 0 or 1), which translates into improved performance metrics. Additionally, looking at the strength of each input species on the resulting prediction \cite{kans,kan2}, shows that the majority of the signal for the American Robin comes from itself. The edge strength of American Robin is $0.68$ compared to the next closest being $0.21$, indicating that LabelKAN is mainly boosting the model's confidence in its original prediction. While the full model is too large to visualize in its entirety, Figure \ref{fig:amerob_kan_plot} shows a subset of the learned KAN structure for the American Robin, clearly highlighting that the majority of the signal comes from the species itself, where the signal strength from the input species is much higher for the American Robin, with only small weights from other species.

By comparison, for the least prevalent species, the Northern Shoveler, the edge strength associated with itself is much lower - only $0.283$. The next highest contributor, the Bonaparte's Gull, has an edge strength of $0.17$, which is much closer to the target species edge score than in the case of the American Robin. As shown in Figure \ref{fig:norsho_kan_plot}, the top 5 contributing species for the Northern Shoveler, looking at the strength of the signal coming from the species inputs, have a more evenly distributed signal reflecting greater importance of co-occurring species.

\begin{figure}[t]
    \centering
    \includegraphics[width=\linewidth]{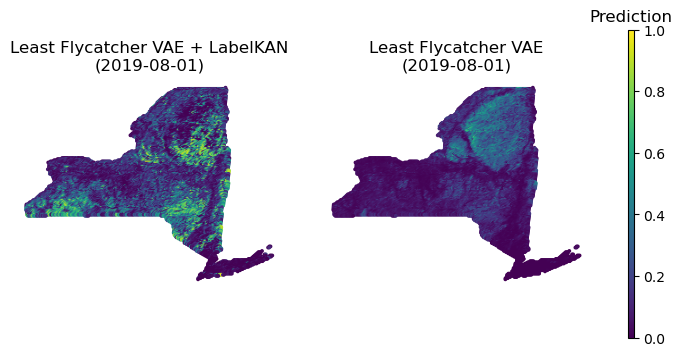}
    \caption{Predictions for the Least Flycatcher, a rare species, on August 1, 2019 for the VAE + LabelKAN (left) and base VAE (right). Despite the KAN only receiving information about the predicted community from the VAE, it significantly improves the species predictions in the Adirondacks and southwestern New York compared to the base model (validation by experts at the Cornell Lab of Ornithology), showing the value of learning from co-occurring species.}
    \label{fig:leafly}
\end{figure}

LabelKAN also improves the spatial patterns of species distributions made by the base model. For example, consider the Least Flycatcher, a relatively uncommon species observed on only $2.6\%$ of checklists. As shown in Figure \ref{fig:leafly}, the base VAE model struggles to differentiate locations within the Adirondacks region and predicting a relatively low probability of presence across the entire range where the Least Flycatcher is present. In contrast, LabelKAN is able to produce predictions that more closely align with known distributions of the Least Flycatcher within the Adirondacks (validation by experts at the Cornell Lab of Ornithology), just by using information from the logit values of species predictions from the base VAE. This suggests that LabelKAN is able to learn important co-occurrence patterns and extract meaningful signals from the presence of other species.

\subsection{Great Blue Heron}

\begin{figure}[t!]
    \centering
    \begin{subfigure}{\linewidth}
        \centering
        \includegraphics[width=\linewidth]{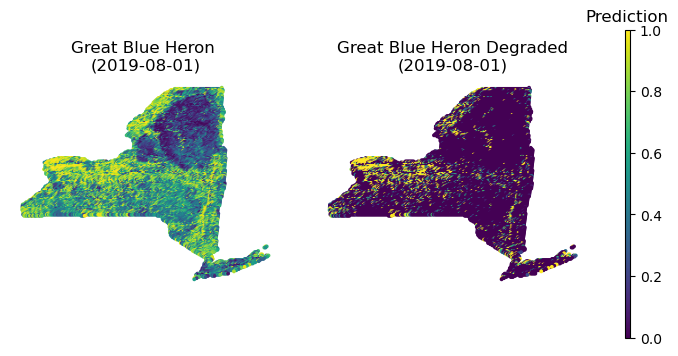}
        \caption{Predicted probability of presence of the Great Blue Heron for August 1, 2019 (left) and the map with forced degradation (right).}
        \label{fig:grbher_map}
    \end{subfigure}
    \begin{subfigure}{\linewidth}
        \centering
        \includegraphics[width=\linewidth]{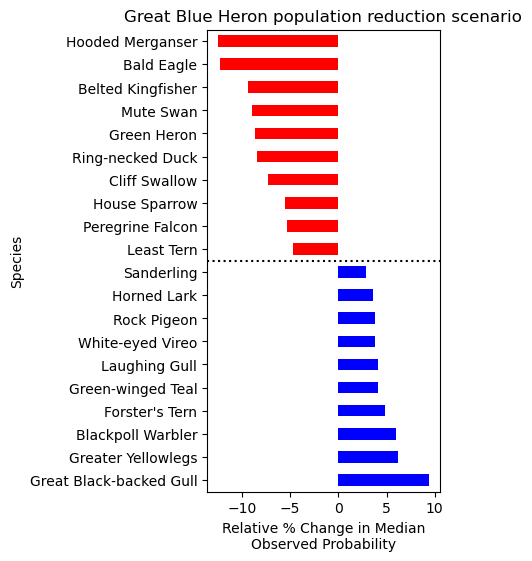}
        \caption{Relative change in median predicted probability due to forced degradation of the Great Blue Heron}
        \label{fig:grbher}
    \end{subfigure}
    \label{fig:grbher_all}
    \caption{Great Blue Heron analysis. Shown are the top 10 species most negatively and positively impacted by changes in Great Blue Heron occurrence. A synthetic $60\%$ decline in occurrence of the Great Blue Heron may be associated with notable impacts on several other water-associated birds.}
\end{figure}

Next, we explore the potential impacts of changing species populations on bird communities by focusing on the Great Blue Heron, an emblematic species undergoing rapid population declines \cite{nadecline,ebirdsnt}.

From the LabelKAN model, we compute predictions for each species $i$ as $y_{\mathcal{K}_i} = \sigma(\mathcal{K}_i(\mathcal{M}(X)))$. Here, the output of $\mathcal{M}(X)$ can be interpreted as learning a species-structured latent space from the environmental input features, where the latent space is constructed from the logits of each species regional in the community. This latent space allows us to synthetically generate initial logits and see the resulting community of species from the altered predictions. To assess the potential effects of a substantial decline in the Great Blue Heron population, we synthetically set its logit value to a large negative number - specifically, the 1st percentile of all logit values for this species - and quantify how predictions for other species change relative to their original values. Although the full joint distribution of the latent space is unknown and sampling out of distribution may lead to poor predictions, we believe that this approach still yields meaningful insights into how other species and communities may be affected. By choosing the manipulated logit within the observed range for the Great Blue Heron, we ensure that the perturbation remains realistic. Original and modified distributions of the Great Blue Heron are shown in Figure \ref{fig:grbher_map}.

First, we construct a $3\times3$km grid across New York state and predict the probability of each species being present at each location on the first of every month during the breeding season (May - August). We then subset these predictions to areas where at least $30\%$ of the surrounding $3\times3$km area is classified as water habitat in the set of environmental features used by the base VAE. From these initial predictions, we extract the logit values produced by the base model for each location and date. To a simulate a decline in the Great Blue Heron, we modify its logit value to be equal to the $1$st percentile of all predicted logits for this species across the validation set. These modified logits are then passed through the LabelKAN heads so that we can compute the relative change in predictions for all other species. 

We present the median relative change in the predicted presence of a subset of species whose prevalence in water habitats exceeds their overall prevalence across New York state. The results, sorted by effect size, are shown in Figure \ref{fig:grbher}. A synthetic decline in the occurrence of the Great Blue Heron by $60\%$ results in notable decreases in several other bird species, including the Bald Eagle, Mute Swan, Belted Kingfisher, and the Green Heron, each showing around a $10\%$ decrease in occurrence. These species align well with the ecological expectations of experts given known co-occurrence patterns. Some of the species that show potential increases in occurrence, such as the Blackpoll Warbler, White-eyed Vireo, and Horned Lark, typically occupy different habitats (forest, scrub, and grasslands, respectively) than the Great Blue Heron and are unlikely to directly co-occur, suggesting that their predicted increases may reflect broader community-level shifts rather than direct associations with the Great Blue Heron.

\section{Discussion}

Our proposed LabelKAN method is able to capture strong species-species co-occurrence patterns beyond what standard joint species distribution models are able to capture. In addition to this, unlike with many standard methods, KANs are able to capture higher order interactions between species within a community as KANs are capable of modeling any function of their inputs, just as with a standard neural network \cite{kans}. We note that while we are able to capture these strong co-occurrence signals, LabelKAN is a correlative model and cannot make causal inference, so truly understanding what species may be impacted by the decline of the Great Blue Heron is difficult to fully capture.

In this work, we focused on separating out the impacts of the community learning on each individual species by constructing a KAN per species. However, because each KAN has as input the entire species initial likelihood predictions, by doing this our methods will naively scale quadratically with the number of species. We leave further improvements for future work, such as learning $k$ representative communities as $k$ KANs that output multiple species at a time.

While most of the work has been focused on presence-absence data, we also see similar improvements with presence-only data and training. We leave further investigation on presence-only data for future work.

\section{Conclusion}

We introduced LabelKAN, a novel framework that applies a series of KANs to the outputs of a base multi-label classification model. LabelKAN leverages the underlying community structure among labels based on the individual likelihoods of each label, independent of the specific base model architecture. Our results demonstrated that the performance gains are not merely a byproduct of the two-stage prediction process. Rather, the KAN itself plays a critical role in improving label predictions, by maintaining species-level information when trying to capture community structure. Incorporating this domain knowledge into the model yields a $2-3\%$ increase in both micro and macro-aggregated AUPRC. LabelKAN effectively learns label-label co-occurrence patterns without greatly compromising performance on any individual species. Moreover, the interpretability of KANs allows for inference on the learned label-label structure. We identified two potential mechanisms for the improved performance on rare and common labels alike, one that enhances rare label predictions by pooling signal from co-occurring labels, and another that boosts the confidence of predictions for more common labels.

This powerful inter-label learning combined with the structured latent space from the base model, allows us to analyze scenarios involving the decline of specific species. By focusing on a careful study of the Great Blue Heron, a species experiencing range-wide population declines, we were able to identify the communities of species that may be most at risk due to their ecological association with it. This approach offers a pathway for targeted conservation efforts aimed at ecologically-associated and co-occurring species.

LabelKAN's modular approach for capturing inter-label connections from a structured latent space makes it applicable to a wide range of multi-label tasks. LabelKAN is able to make stronger inferences about the effects between co-occurring labels; and its shallow and more interpretable KANs improve understanding of the strength and potential functional forms of the effects. We hope LabelKAN will inspire further work across diverse domains.

\section{Acknowledgments}

This project is partially supported by  an AI2050 Senior Fellowship, a Schmidt Sciences program;the National Science Foundation (NSF); the National Institute of Food and Agriculture (USDA/NIFA); the Air Force Office of Scientific Research (AFOSR); the Wolf Creek Foundation; the Advanced Cyberinfrastructure Coordination Ecosystem: Services \& Support (ACCESS) program; and the Cornell Lab of Ornithology.

\bibliography{aaai2026}

\begin{thebibliography}{37}
\providecommand{\natexlab}[1]{#1}

\bibitem[{Amatulli et~al.(2018)Amatulli, Domisch, Tuanmu, Parmentier, Ranipeta, Malczyk, and Jetz}]{topo2}
Amatulli, G.; Domisch, S.; Tuanmu, M.-N.; Parmentier, B.; Ranipeta, A.; Malczyk, J.; and Jetz, W. 2018.
\newblock A suite of global, cross-scale topographic variables for environmental and biodiversity modeling.
\newblock \emph{Scientific Data}, 5(1): 180040.

\bibitem[{Bai, Kong, and Gomes(2022)}]{gmvae}
Bai, J.; Kong, S.; and Gomes, C.~P. 2022.
\newblock Gaussian Mixture Variational Autoencoder with Contrastive Learning for Multi-Label Classification.
\newblock arXiv:2112.00976.

\bibitem[{Becker et~al.(2009)Becker, Sandwell, Smith, Braud, Binder, Depner, Fabre, Factor, Ingalls, Kim, Ladner, Marks, Nelson, Pharaoh, Trimmer, Rosenberg, Wallace, and Weatherall}]{topo1}
Becker, J.~J.; Sandwell, D.~T.; Smith, W. H.~F.; Braud, J.; Binder, B.; Depner, J.; Fabre, D.; Factor, J.; Ingalls, S.; Kim, S.-H.; Ladner, R.; Marks, K.; Nelson, S.; Pharaoh, A.; Trimmer, R.; Rosenberg, J.~V.; Wallace, G.; and Weatherall, P. 2009.
\newblock Global Bathymetry and Elevation Data at 30 Arc Seconds Resolution: SRTM30\_PLUS.
\newblock \emph{Marine Geodesy}, 32(4): 355--371.

\bibitem[{Boutell et~al.(2004)Boutell, Luo, Shen, and Brown}]{binary_relevance}
Boutell, M.~R.; Luo, J.; Shen, X.; and Brown, C.~M. 2004.
\newblock Learning multi-label scene classification.
\newblock \emph{Pattern Recognition}, 37(9): 1757--1771.

\bibitem[{Carroll et~al.(2017)Carroll, DiMiceli, Wooten, Hubbard, Sohlberg, and Townshend}]{modis}
Carroll, M.~L.; DiMiceli, C.~M.; Wooten, M.~R.; Hubbard, A.~B.; Sohlberg, R.~A.; and Townshend, J. R.~G. 2017.
\newblock Mod44w Modis/Terra Land Water Mask Derived from Modis and Srtm 13 Global 250m Sin Grid v006.

\bibitem[{Chen et~al.(2019)Chen, Bai, Zhao, Ament, Gregoire, and Gomes}]{drnets}
Chen, D.; Bai, Y.; Zhao, W.; Ament, S.; Gregoire, J.~M.; and Gomes, C.~P. 2019.
\newblock Deep Reasoning Networks: Thinking Fast and Slow.
\newblock arXiv:1906.00855.

\bibitem[{Chen, Xue, and Gomes(2018)}]{dmvp}
Chen, D.; Xue, Y.; and Gomes, C.~P. 2018.
\newblock End-to-End Learning for the Deep Multivariate Probit Model.
\newblock arXiv:1803.08591.

\bibitem[{Cole et~al.(2023)Cole, Horn, Lange, Shepard, Leary, Perona, Loarie, and Aodha}]{SINR}
Cole, E.; Horn, G.~V.; Lange, C.; Shepard, A.; Leary, P.; Perona, P.; Loarie, S.; and Aodha, O.~M. 2023.
\newblock Spatial Implicit Neural Representations for Global-Scale Species Mapping.
\newblock arXiv:2306.02564.

\bibitem[{Davis et~al.(2023)Davis, Bai, Chen, Robinson, Ruiz-Gutierrez, Gomes, and Fink}]{dmvp2}
Davis, C.~L.; Bai, Y.; Chen, D.; Robinson, O.; Ruiz-Gutierrez, V.; Gomes, C.~P.; and Fink, D. 2023.
\newblock Deep learning with citizen science data enables estimation of species diversity and composition at continental extents.
\newblock \emph{Ecology}, 104(12): e4175.

\bibitem[{Elith and Leathwick(2009)}]{sdms}
Elith, J.; and Leathwick, J.~R. 2009.
\newblock Species Distribution Models: Ecological Explanation and Prediction Across Space and Time.
\newblock \emph{Annual Review of Ecology, Evolution, and Systematics}, 40(Volume 40, 2009): 677--697.

\bibitem[{Fink et~al.(2023)Fink, Auer, Johnston, Strimas-Mackey, Ligocki, Robinson, Hochachka, Jaromczyk, Crowley, Dunham, Stillman, Davies, Rodewald, Ruiz-Gutierrez, and Wood}]{ebirdsnt}
Fink, D.; Auer, T.; Johnston, A.; Strimas-Mackey, M.; Ligocki, S.; Robinson, O.; Hochachka, W.; Jaromczyk, L.; Crowley, C.; Dunham, K.; Stillman, A.; Davies, I.; Rodewald, A.; Ruiz-Gutierrez, V.; and Wood, C. 2023.
\newblock eBird Status and Trends, Data Version: 2022.

\bibitem[{Guillera-Arroita(2017)}]{sdmsimperfect}
Guillera-Arroita, G. 2017.
\newblock Modelling of species distributions, range dynamics and communities under imperfect detection: advances, challenges and opportunities.
\newblock \emph{Ecography}, 40(2): 281--295.

\bibitem[{Guisan, Edwards, and Hastie(2002)}]{glmgam}
Guisan, A.; Edwards, T.~C.; and Hastie, T. 2002.
\newblock Generalized linear and generalized additive models in studies of species distributions: setting the scene.
\newblock \emph{Ecological Modelling}, 157(2): 89--100.

\bibitem[{Guisan and Thuiller(2005)}]{Guisan}
Guisan, A.; and Thuiller, W. 2005.
\newblock Predicting species distribution: offering more than simple habitat models.
\newblock \emph{Ecology Letters}, 8(9): 993--1009.

\bibitem[{Guisan and Zimmermann(2000)}]{GUISAN2000147}
Guisan, A.; and Zimmermann, N.~E. 2000.
\newblock Predictive habitat distribution models in ecology.
\newblock \emph{Ecological Modelling}, 135(2): 147--186.

\bibitem[{Hu, Si-Moussi, and Thuiller(2024)}]{deepsdms}
Hu, Y.; Si-Moussi, S.; and Thuiller, W. 2024.
\newblock Introduction to deep learning methods for multi-species predictions.
\newblock \emph{Methods in Ecology and Evolution}.

\bibitem[{iNaturalist()}]{inat}
iNaturalist. 2023.
\newblock iNaturalist.
\newblock Available from https://www.inaturalist.org. Accessed: May 9, 2023.

\bibitem[{Jin and Nakayama(2016)}]{annotationordermatters}
Jin, J.; and Nakayama, H. 2016.
\newblock Annotation Order Matters: Recurrent Image Annotator for Arbitrary Length Image Tagging.
\newblock arXiv:1604.05225.

\bibitem[{Johnston et~al.(2025)Johnston, Rodewald, Strimas-Mackey, Auer, Hochachka, Stillman, Davis, Ruiz-Gutierrez, Dokter, Miller, Robinson, Ligocki, Jaromczyk, Crowley, Wood, and Fink}]{nadecline}
Johnston, A.; Rodewald, A.~D.; Strimas-Mackey, M.; Auer, T.; Hochachka, W.~M.; Stillman, A.~N.; Davis, C.~L.; Ruiz-Gutierrez, V.; Dokter, A.~M.; Miller, E.~T.; Robinson, O.; Ligocki, S.; Jaromczyk, L.~O.; Crowley, C.; Wood, C.~L.; and Fink, D. 2025.
\newblock North American bird declines are greatest where species are most abundant.
\newblock \emph{Science}, 388(6746): 532--537.

\bibitem[{Liu et~al.(2024)Liu, Ma, Wang, Matusik, and Tegmark}]{kan2}
Liu, Z.; Ma, P.; Wang, Y.; Matusik, W.; and Tegmark, M. 2024.
\newblock KAN 2.0: Kolmogorov-Arnold Networks Meet Science.
\newblock arXiv:2408.10205.

\bibitem[{Liu et~al.(2025)Liu, Wang, Vaidya, Ruehle, Halverson, Soljačić, Hou, and Tegmark}]{kans}
Liu, Z.; Wang, Y.; Vaidya, S.; Ruehle, F.; Halverson, J.; Soljačić, M.; Hou, T.~Y.; and Tegmark, M. 2025.
\newblock KAN: Kolmogorov-Arnold Networks.
\newblock arXiv:2404.19756.

\bibitem[{Loshchilov and Hutter(2019)}]{adamw}
Loshchilov, I.; and Hutter, F. 2019.
\newblock Decoupled Weight Decay Regularization.
\newblock arXiv:1711.05101.

\bibitem[{McDermott et~al.(2025)McDermott, Zhang, Hansen, Angelotti, and Gallifant}]{auprcauroc}
McDermott, M. B.~A.; Zhang, H.; Hansen, L.~H.; Angelotti, G.; and Gallifant, J. 2025.
\newblock A Closer Look at AUROC and AUPRC under Class Imbalance.
\newblock arXiv:2401.06091.

\bibitem[{Ovaskainen and Soininen(2011)}]{hierarchicalsparse}
Ovaskainen, O.; and Soininen, J. 2011.
\newblock Making more out of sparse data: hierarchical modeling of species communities.
\newblock \emph{Ecology}, 92(2): 289--295.

\bibitem[{Paszke et~al.(2019)Paszke, Gross, Massa, Lerer, Bradbury, Chanan, Killeen, Lin, Gimelshein, Antiga, Desmaison, Köpf, Yang, DeVito, Raison, Tejani, Chilamkurthy, Steiner, Fang, Bai, and Chintala}]{pytorch}
Paszke, A.; Gross, S.; Massa, F.; Lerer, A.; Bradbury, J.; Chanan, G.; Killeen, T.; Lin, Z.; Gimelshein, N.; Antiga, L.; Desmaison, A.; Köpf, A.; Yang, E.; DeVito, Z.; Raison, M.; Tejani, A.; Chilamkurthy, S.; Steiner, B.; Fang, L.; Bai, J.; and Chintala, S. 2019.
\newblock PyTorch: An Imperative Style, High-Performance Deep Learning Library.
\newblock arXiv:1912.01703.

\bibitem[{Pollock et~al.(2014)Pollock, Tingley, Morris, Golding, O'Hara, Parris, Vesk, and McCarthy}]{jsdms}
Pollock, L.~J.; Tingley, R.; Morris, W.~K.; Golding, N.; O'Hara, R.~B.; Parris, K.~M.; Vesk, P.~A.; and McCarthy, M.~A. 2014.
\newblock Understanding co-occurrence by modelling species simultaneously with a Joint Species Distribution Model (JSDM).
\newblock \emph{Methods in Ecology and Evolution}, 5(5): 397--406.

\bibitem[{Read et~al.(2011)Read, Pfahringer, Holmes, and Frank}]{classifier_chains}
Read, J.; Pfahringer, B.; Holmes, G.; and Frank, E. 2011.
\newblock Classifier chains for multi-label classification.
\newblock \emph{Machine Learning}, 85(3): 333--359.

\bibitem[{Ryckewaert et~al.(2025)Ryckewaert, Marcos, Botella, Servajean, Bonnet, and Joly}]{deepmaxent}
Ryckewaert, M.; Marcos, D.; Botella, C.; Servajean, M.; Bonnet, P.; and Joly, A. 2025.
\newblock Applying the maximum entropy principle to neural networks enhances multi-species distribution models.
\newblock arXiv:2412.19217.

\bibitem[{Ryo et~al.(2021)Ryo, Angelov, Mammola, Kass, Benito, and Hartig}]{eai_eco}
Ryo, M.; Angelov, B.; Mammola, S.; Kass, J.~M.; Benito, B.~M.; and Hartig, F. 2021.
\newblock Explainable artificial intelligence enhances the ecological interpretability of black-box species distribution models.
\newblock \emph{Ecography}, 44(2): 199--205.

\bibitem[{{Secretariat of the Convention on Biological Diversity}(2022)}]{cbd}
{Secretariat of the Convention on Biological Diversity}. 2022.
\newblock Convention on Biological Diversity.
\newblock Accessed: 2025-02-03.

\bibitem[{Sullivan et~al.(2014)Sullivan, Aycrigg, Barry, Bonney, Bruns, Cooper, Damoulas, Dhondt, Dietterich, Farnsworth, Fink, Fitzpatrick, Fredericks, Gerbracht, Gomes, Hochachka, Iliff, Lagoze, {La Sorte}, Merrifield, Morris, Phillips, Reynolds, Rodewald, Rosenberg, Trautmann, Wiggins, Winkler, Wong, Wood, Yu, and Kelling}]{ebird}
Sullivan, B.~L.; Aycrigg, J.~L.; Barry, J.~H.; Bonney, R.~E.; Bruns, N.; Cooper, C.~B.; Damoulas, T.; Dhondt, A.~A.; Dietterich, T.; Farnsworth, A.; Fink, D.; Fitzpatrick, J.~W.; Fredericks, T.; Gerbracht, J.; Gomes, C.; Hochachka, W.~M.; Iliff, M.~J.; Lagoze, C.; {La Sorte}, F.~A.; Merrifield, M.; Morris, W.; Phillips, T.~B.; Reynolds, M.; Rodewald, A.~D.; Rosenberg, K.~V.; Trautmann, N.~M.; Wiggins, A.; Winkler, D.~W.; Wong, W.-K.; Wood, C.~L.; Yu, J.; and Kelling, S. 2014.
\newblock The eBird enterprise: An integrated approach to development and application of citizen science.
\newblock \emph{Biological Conservation}, 169: 31--40.

\bibitem[{Thornton et~al.(2022)Thornton, Shrestha, Wei, Thornton, and Kao}]{daymet}
Thornton, M.; Shrestha, R.; Wei, Y.; Thornton, P.; and Kao, S.-C. 2022.
\newblock Daymet: Daily Surface Weather Data on a 1-km Grid for North America, Version 4 R1.

\bibitem[{Tsoumakas, Katakis, and Vlahavas(2010)}]{label_powerset}
Tsoumakas, G.; Katakis, I.; and Vlahavas, I. 2010.
\newblock \emph{Mining Multi-label Data}, 667--685.
\newblock Boston, MA: Springer US.
\newblock ISBN 978-0-387-09823-4.

\bibitem[{{United Nations General Assembly}(2015)}]{sdgs}
{United Nations General Assembly}. 2015.
\newblock Transforming our world: the 2030 Agenda for Sustainable Development.
\newblock Accessed: 2022-11-23.

\bibitem[{Wang et~al.(2016)Wang, Yang, Mao, Huang, Huang, and Xu}]{cnnrnn}
Wang, J.; Yang, Y.; Mao, J.; Huang, Z.; Huang, C.; and Xu, W. 2016.
\newblock CNN-RNN: A Unified Framework for Multi-label Image Classification.
\newblock arXiv:1604.04573.

\bibitem[{Wisz et~al.(2012)Wisz, Pottier, Kissling, Pellissier, Lenoir, Damgaard, Dormann, Forchhammer, Grytnes, Guisan, Heikkinen, H{\o}ye, K{\"u}hn, Luoto, Maiorano, Nilsson, Normand, {\"O}ckinger, Schmidt, Termansen, Timmermann, Wardle, Aastrup, and Svenning}]{Wisz2012}
Wisz, M.~S.; Pottier, J.; Kissling, W.~D.; Pellissier, L.; Lenoir, J.; Damgaard, C.~F.; Dormann, C.~F.; Forchhammer, M.~C.; Grytnes, J.-A.; Guisan, A.; Heikkinen, R.~K.; H{\o}ye, T.~T.; K{\"u}hn, I.; Luoto, M.; Maiorano, L.; Nilsson, M.-C.; Normand, S.; {\"O}ckinger, E.; Schmidt, N.~M.; Termansen, M.; Timmermann, A.; Wardle, D.~A.; Aastrup, P.; and Svenning, J.-C. 2012.
\newblock The role of biotic interactions in shaping distributions and realised assemblages of species: implications for species distribution modelling.
\newblock \emph{Biol Rev Camb Philos Soc}, 88(1): 15--30.

\bibitem[{Yuan et~al.(2024)Yuan, Chen, Zhang, Shi, Geng, Fan, and Rui}]{graphattention}
Yuan, J.; Chen, S.; Zhang, Y.; Shi, Z.; Geng, X.; Fan, J.; and Rui, Y. 2024.
\newblock Graph Attention Transformer Network for Multi-Label Image Classification.
\newblock arXiv:2203.04049.

\end{thebibliography}


\end{document}